\documentclass[12pt]{article}

\usepackage{amssymb,amsmath}

 \catcode`\@=11 \@addtoreset{equation}{section}\catcode`\@=12
\newcommand{\R}{ {\mathbb R} }

\begin{document}

 \begin{center}

 \large \bf Examples of stable  exponential cosmological  solutions with three factor spaces in EGB  model  with a $\Lambda$-term 
  \end{center}

 \vspace{0.3truecm}

 \begin{center}

   K. K. Ernazarov$^{1}$ and V. D. Ivashchuk$^{1,2}$ 

\vspace{0.3truecm}

 \it $^{1}$Institute of Gravitation and Cosmology,
 RUDN University, 6 Miklukho-Maklaya ul.,
 Moscow 117198, Russia

 \it $^{2}$ Center for Gravitation and Fundamental Metrology,
 VNIIMS, 46 Ozyornaya ul., Moscow 119361, Russia

\end{center}

\begin{abstract}

We deal with Einstein-Gauss-Bonnet model in dimension $D$ with a $\Lambda$-term.
We obtain   three stable cosmological solutions with exponential behavior (in time) of three scale factors
corresponding to subspaces of dimensions $(l_0,l_1,l_2) =  (3, 4, 4), (3, 3, 2), (3, 4, 3)$ and $D = 12, 9, 11$, respectively. Any solution  may describe an exponential expansion of $3$-dimensional subspace governed by Hubble parameter $H$. Two of them may also describe a small enough variation of the effective gravitational constant $G$ (in Jordan frame) for certain values of $\Lambda$. 

\end{abstract}


\section{Introduction}

Here we study a $D$-dimensional  Einstein-Gauss-Bonnet (EGB)  gravitational model which contains  Gauss-Bonnet term and  $\Lambda$-term. As it is well-known, the  Gauss-Bonnet term appeared in (super)string theory in the next to leading order correction (in  slope parameter) to the effective action \cite{Zwiebach}-\cite{GW}. Currently, the EGB  model and its extensions, see \cite{Ishihara}-\cite{ChTop-17} and references therein, are under a wide studying in cosmology  aimed at possible  explanation of accelerating expansion of the Universe (i.e. in a context of the so-called dark energy problem) \cite{Riess, Perl, Kowalski}. 
Here we  restrict ourselves by non-singular cosmological solutions with  three scale factors 
$a_i = \exp(H_i t)$, $i =0,1, 2$, where $t$ is a (synchronous) time variable. We find three exact solutions with three different Hubble-like parameters $H_0 >0$, $H_1$ and $H_2$, which correspond to three subspaces of dimensions $3$, $l_1$ and $l_2$ with $(l_1, l_2) = (4, 4), (3, 2), (4, 3)$. Here we put $H_0 =H$, where $H$ is the Hubble parameter \cite{Ade}. Due to results of ref. \cite{Ivas-16} (see also \cite{Pavl-15,ErIvKob-16}), the solutions are  stable in a class of cosmological solutions with diagonal metrics.  It is shown that two obtained solutions may describe a small enough   variation of the effective gravitational constant $G$ (in Jordan frame) \cite{RZ-98, I-96, BIM, Mel} when cosmological constant $\Lambda$ obeys certain restrictions.

\section{The cosmological model}

The action of the model reads
\begin{equation}
  S =  \int_{M} d^{D}z \sqrt{|g|} \{ \alpha_1 (R[g] - 2 \Lambda) +
              \alpha_2 {\cal L}_2[g] \}.
 \label{2.0}
\end{equation}
Here $g = g_{MN} dz^{M} \otimes dz^{N}$ is the metric defined on
the manifold $M$, ${\dim M} = D$, $|g| = |\det (g_{MN})|$, $\Lambda$ is
the cosmological term, $R[g]$ is scalar curvature,
$${\cal L}_2[g] = R_{MNPQ} R^{MNPQ} - 4 R_{MN} R^{MN} +R^2$$
is the  Gauss-Bonnet term and  $\alpha_1$, $\alpha_2$ are
nonzero constants.

We consider the manifold
\begin{equation}
   M = \R  \times   M_1 \times \ldots \times M_n 
   \label{2.1}
\end{equation}
with the metric
\begin{equation}
   g= - d t \otimes d t  +
      \sum_{i=1}^{n} B_i e^{2v^i t} dy^i \otimes dy^i,
  \label{2.2}
\end{equation}
where   $B_i > 0$ are  constants, $i = 1, \dots, n$, and
$M_1, \dots,  M_n$  are one-dimensional manifolds (either compact or non-compact ones)
and $n > 3$.

The equations of motion for the action (\ref{2.0}) 
read \cite{ErIvKob-16}
\begin{eqnarray}
  E = G_{ij} v^i v^j + 2 \Lambda
  - \alpha   G_{ijkl} v^i v^j v^k v^l = 0,  \label{2.3} \\
   Y_i =  \left[ 2   G_{ij} v^j
    - \frac{4}{3} \alpha  G_{ijkl}  v^j v^k v^l \right] \sum_{i=1}^n v^i 
    - \frac{2}{3}   G_{ij} v^i v^j  +  \frac{8}{3} \Lambda = 0,
   \label{2.4}
\end{eqnarray}
$i = 1,\ldots, n$, where  $\alpha = \alpha_2/\alpha_1$. Here \cite{Iv-09,Iv-10}
\begin{equation}
 G_{ij} = \delta_{ij} -1, \qquad   G_{ijkl}  = G_{ij} G_{ik} G_{il} G_{jk} G_{jl} G_{kl}.
\label{2.4G}
\end{equation}

\section{Solutions with three factor spaces}

In this section we present a class of solutions to the set of equations (\ref{2.3}), 
(\ref{2.4}) of the following form:
\begin{equation}
  \label{3.1}
   v =(\overbrace{h_0, \ldots, h_0}^{k_0}, 
   \overbrace{h_1, \ldots, h_1}^{k_1}, \overbrace{h_2, \ldots, h_2}^{k_2}),
\end{equation}
where $h_0$ is the Hubble-like parameter corresponding  
to an $k_0$-dimensional factor space with $k_0 > 1$,  $h_1$ is the Hubble-like parameter 
corresponding to a $k$-dimensional factor space and $h_2$ ($h_2 \neq h_1$) is the Hubble-like parameter 
corresponding to another $k$-dimensional factor space,  $k > 1$.

We consider the ansatz (\ref{3.1}) with three Hubble parameters $H$, $h_1$ and $h_2$ 
which obey the following restrictions:
   \begin{equation}
     h_0 \neq h_1, \quad  h_0 \neq h_2, \quad  h_1 \neq h_2, \quad S_1 = k_0 h_0 + k ( h_1 +  h_2) \neq 0.
   \label{3.3}
   \end{equation}
  
  We put 
  \begin{equation}
       \alpha   >  0.       \label{3.10c}
  \end{equation}

Here we present three examples of  solutions  to equations (\ref{2.3}), (\ref{2.4}) 
with the restrictions (\ref{3.3}) imposed. 
These solutions are obtained by using Mathematica. Any solution 
contains at least one number $3$ among $k_0, k_1, k_2$. 
We renumerate the  subspaces in such way that renumerated 
set $\bar{v}$ has the form

\begin{equation}
  \label{3.1r}
  \bar{v} =(\underbrace{H, H, H}_{l_0 = 3}, 
   \underbrace{H_1, \ldots, H_1}_{l_1}, \underbrace{H_2, \ldots, H_2}_{l_2}),
\end{equation}
$H > 0$ and the metric (\ref{2.2}) reads
\begin{equation}
   g= - d t \otimes d t  +
      \sum_{i=1}^{n} \bar{B}_i e^{2 \bar{v}^i t} d\bar{y}^i \otimes d\bar{y}^i,
  \label{2.2r}
\end{equation}
where  $\bar{v}^i = v^{\sigma(i)}$, $\bar{B}_i = B_{\sigma(i)} > 0$ and 
$\bar{y}^i = y^{\sigma(i)} \equiv x^i$, $i = 1, \dots, n$, for certain permutation $\sigma \in S_n$.

The  metric (\ref{2.2r}) may be rewritten as 
\begin{eqnarray}
   g=  - d t \otimes d t  +  e^{2 H t} \sum_{i=1}^{3} dx^i \otimes dx^i
      \nonumber \\ 
      + e^{2 H_1 t} \sum_{i=4}^{3+ l_1} dx^i \otimes dx^i + 
      e^{2 H_2 t} \sum_{i=4 + l_1}^{3+ l_1 + l_2} dx^i \otimes dx^i,
  \label{2.2s}
\end{eqnarray} 
where $H > 0$ is Hubble parameter and the set of parameters $(H,H_1,H_2)$ 
 is obtained from the set $(h_0,h_1,h_2)$ by certain permutation.

The dimensionless parameter of variation of (effective) gravitational constant
(in Jordan frame) \cite{RZ-98,I-96,BIM,Mel} reads

\begin{equation}
  \label{3.var}
{\rm var} = \frac{\dot{G}}{G H} = - \frac{l_1 H_1 + l_2 H_2}{H}. 
\end{equation}

Due to the experimental data, the variation of the gravitational constant is allowed
at the level of $10^{-13}$ per year and less.
Here one may  use   the following constraint on the magnitude of the dimensionless variation of the
 effective gravitational constant:

 \begin{equation}
 \label{5.G1}
  - 0,63 \cdot 10^{-3} < \frac{\dot{G}}{GH} < 1,13 \cdot 10^{-3}.
 \end{equation}
  It comes from the most stringent limitation
  on $G$-dot obtained by the set of ephemerides \cite{Pitjeva}

\begin{equation}
 \label{5.G2}
          -0,42 \cdot 10^{-13} \ yr^{-1} <  \dot{G}/G  <  0,75  \cdot 10^{-13} \ yr^{-1}
\end{equation}
allowed at 95\% confidence (2$\sigma$) level
and the present value of the Hubble parameter \cite{Ade}
\begin{equation}
 \label{H}
  H_0 =  (67,3 \pm 2,4) \ km/s \ Mpc^{-1} =  (6,878  \pm 0,245) \cdot 10^{-11} \ yr^{-1},
 \end{equation}
  with 95\% confidence level ($2 \sigma$).

In what follows  we denote $\lambda = \Lambda \alpha$, $\alpha >0$. 

\subsection{The case $k_0 =3$, $k = 4$ }

 Let $k_0 =3$, $k =4$. Here we put $k_0 = l_0 =3$, $l_1 = k_1 = 4$, $l_2 = k_2 =4$
 and  $(l_0,l_1,l_2) = (3,4,4)$.
 The solution reads: 

\begin{equation}
H = h_0 = \frac{\sqrt{15}}{20 \sqrt{\alpha}} \sqrt{1 + \sqrt{480\lambda -79}} > 0,
   \label{6.1A}
 \end{equation}

\begin{equation}
H_1 = h_1 = - \frac{\sqrt{15}}{60\sqrt{\alpha}} 
   \Biggl(\sqrt{1 + \sqrt{480\lambda -79}}
      - 2\sqrt{4 -  \sqrt{480\lambda -79}}\Biggr),
   \label{6.2A}
 \end{equation}

\begin{equation}
H_2 = h_2 =- \frac{\sqrt{15}}{60 \sqrt{\alpha}} \Biggl(\sqrt{1 + \sqrt{480\lambda -79}} 
+ 2\sqrt{4 -  \sqrt{480\lambda -79}}\Biggr).
   \label{6.3A}
 \end{equation}

\begin{equation}
S_1 = \frac{\sqrt{15}}{60\sqrt{\alpha}}\sqrt{1 + \sqrt{480\lambda -79}}.
   \label{6.4A}
 \end{equation}

Here 
\begin{equation}
     \frac{79}{480} < \lambda < \frac{19}{96}.
   \label{6.4L}
 \end{equation}

We obtain a huge value for the dimensionless variation of $G$
\begin{equation}
  \label{3.var1}
  {\rm var} = 8/3, 
\end{equation}
which does not obey restriction (\ref{5.G1}).  

\subsection{The case $ k_0 =2$, $k = 3$ }

Let us put  $k_0 =2$, $k=3$.  Here $l_0 = k_2 = 3$, $l_1 = k_1 = 3$, $l_2 = k_0 = 2$, 
and  $(l_0,l_1,l_2) = (3,3,2)$. The solution reads:

\begin{equation}
H = h_2 =- \frac{\sqrt{15}}{60\sqrt{\alpha}}\Biggl(\sqrt{1+ 2\sqrt{60\lambda - 11}}
  -  \sqrt{5} \sqrt{5 -  2\sqrt{60\lambda - 11}}\Biggr) > 0,
   \label{6.5A}
 \end{equation}

\begin{equation}
H_1 = h_1 = - \frac{\sqrt{15}}{60\sqrt{\alpha}} \Biggl(\sqrt{1+ 2\sqrt{60\lambda - 11}}
  + \sqrt{5} \sqrt{5 -  2\sqrt{60\lambda - 11}} \Biggr),   \label{6.6A}
 \end{equation}

\begin{equation}
H_2 = h_0 = \frac{1}{\sqrt{15 \sqrt{\alpha}}}\sqrt{1+ 2\sqrt{60\lambda -11}},
   \label{6.7A}
 \end{equation}

\begin{equation}
S_1 = \frac{\sqrt{15}}{30 \sqrt{\alpha}}\sqrt{1+ 2\sqrt{60\lambda -11}}.   
\label{6.8A}
 \end{equation}

Here  we have the following restriction on $\lambda$
\begin{equation}
     \frac{11}{60} < \lambda < \frac{2209}{8640}.
   \label{6.8L}
 \end{equation}

For $\lambda = \lambda_1 = 213/980$ we get
$H = \frac{1}{2\sqrt{35 \alpha}}$, $H_1 = - 4H$, $H_2 = 6 H$, $S_1 = 3H$, ${\rm var} = 0$ 
in agreement with \cite{ErIv-17-2}.

In case when  the following restriction is imposed:   
 \begin{equation}
  |{\rm var}| < \delta 
  \label{6.8var}
 \end{equation} 
there exists $\epsilon_1 > 0$ such that 
the (\ref{6.8var}) is obeyed if   
\begin{equation}
  | \lambda- \lambda_1| < \epsilon_1.
  \label{6.8L1}
 \end{equation} 
This follows from the continuity of the function ${\rm var} = {\rm var}(\lambda)$.

\subsection{The case $ k_0 =4$, $k = 3$ }

Now we put  $k_0 =4$, $k=3$ and  $l_0 = k_1 = 3$, $l_1 = k_0 = 4$, $l_2 = k_2 = 3$,
 and  $(l_0,l_1,l_2) = (3,4,3)$. The solution reads:

\begin{equation}
H  = h_1 =  \frac{\sqrt{5}}{140 \sqrt{\alpha}}\Biggl(\sqrt{21} \sqrt{3+ 2\sqrt{81 - 420\lambda}} 
-  7\sqrt{7 -  2\sqrt{81 - 420\lambda}}\Biggr),
   \label{6.9A}
 \end{equation}

\begin{equation}
H_1 = h_0 =- \frac{1}{\sqrt{105\alpha}}\sqrt{3+ 2\sqrt{81 - 420\lambda}},  
   \label{6.10A}
 \end{equation}

\begin{equation}
H_2 = h_2 = \frac{\sqrt{5}}{140 \sqrt{\alpha}}\Biggl(\sqrt{21} \sqrt{3+ 2\sqrt{81 - 420\lambda}}
 +  7\sqrt{7 -  2\sqrt{81 - 420\lambda}}\Biggr),
   \label{6.11A}
 \end{equation}

\begin{equation}
S_1 = \frac{\sqrt{105}}{210\sqrt{\alpha}}\sqrt{3+ 2\sqrt{81 - 420\lambda}}.  
   \label{6.10S}
 \end{equation}

Here  we impose
\begin{equation}
     \frac{55}{336} < \lambda < \frac{11}{60}.
   \label{6.10L}
 \end{equation}

For $\lambda = \lambda_2 =  1443/8092$ we get
$H = \frac{1}{2\sqrt{119 \alpha}}$, $H_1 = - 6H$, $H_2 = 8H$, $S_1 = 3H$ and ${\rm var} = 0$
in agreement with \cite{ErIv-17-2}.

When the  restriction  (\ref{6.8var}) is imposed  
  there exists $\epsilon_2 > 0$ such that 
the (\ref{6.8var}) is obeyed if   
\begin{equation}
  | \lambda- \lambda_2| < \epsilon_2.
  \label{6.8L1}
 \end{equation}

\section{The analysis of stability}

The solutions obey two restrictions:
\begin{equation}
  S_1 = \sum_{i = 1}^{n} v^i  >0.
 \label{4.1}
\end{equation}
and 
\begin{equation}
  \det (L_{ij}(v)) \neq 0,
  \label{4.2}
\end{equation}
where 
\begin{equation}
L =(L_{ij}(v)) = (2 G_{ij} - 4 \alpha G_{ijks} v^k v^s).
   \label{4.1b}
 \end{equation}
 Relation (\ref{4.2}) was proved in ref.  \cite{ErIv-17-2} by using 
 restrictions from (\ref{3.3}).

We remind that for general cosmological setup with the metric 
\begin{equation}
 g= - dt \otimes dt + \sum_{i=1}^{n} e^{2\beta^i(t)}  dy^i \otimes dy^i,
 \label{4.3}
\end{equation}
we have  the  set of  equations \cite{ErIvKob-16} 
\begin{eqnarray}
     E = G_{ij} h^i h^j + 2 \Lambda  - \alpha G_{ijkl} h^i h^j h^k h^l = 0,
         \label{4.3.1} \\
         Y_i =  \frac{d L_i}{dt}  +  (\sum_{j=1}^n h^j) L_i -
                 \frac{2}{3} (G_{sj} h^s h^j -  4 \Lambda) = 0,
                     \label{4.3.2a}
          \end{eqnarray}
where $h^i = \dot{\beta}^i$,           
 \begin{equation}
  L_i = L_i(h) = 2  G_{ij} h^j
       - \frac{4}{3} \alpha  G_{ijkl}  h^j h^k h^l  
       \label{4.3.3},
 \end{equation}
 $i = 1,\ldots, n$.

Due to results of ref. \cite{Ivas-16}  a fixed point solution
$(h^i(t)) = (v^i)$ ($i = 1, \dots, n$; $n >3$) to eqs. (\ref{4.3.1}), (\ref{4.3.2a})
obeying restrictions (\ref{4.1}) and   (\ref{4.2}) is  stable under perturbations
\begin{equation}
 h^i(t) = v^i +  \delta h^i(t), 
\label{4.3h}
\end{equation}
 $i = 1,\ldots, n$,  as $t \to + \infty$.
This follows from the relations   \cite{Ivas-16}
      \begin{equation}
          \delta h^i = A^i \exp( - S_1(v) t ),   \qquad        
            \sum_{i =1}^{n} C_i(v)  A^i =0,
            \label{4.16}
      \end{equation}
       ($A^i$ are constants) $i = 1, \dots, n$, 
which are valid when restrictions (\ref{4.1}), (\ref{4.2}) are imposed.   
Thus,  all solutions under consideration  are stable.

\section{Conclusions}
Here we were studying the  Einstein-Gauss-Bonnet (EGB) model in 
$D$ dimensions  with the cosmological term $\Lambda$. We have obtained three examples of non-singular solutions with exponential time dependence (with respect to  synchronous time variable $t$) of three scale factors, governed by 
three non-coinciding Hubble-like parameters: $H_0 >0$, $
H_1$ and $H_2$, which correspond to factor spaces of dimensions $3$, $l_1$ and $l_2$, respectively, and $D = 4 + l_1 + l_2 $. Here $(l_1, l_2) = (4, 4), (3, 2), (4, 3)$. Due to  results of ref. \cite{Ivas-16}  all the obtained solutions are stable (as $t \to + \infty$). We put $H_0 = H$, where $H$ is Hubble parameter. We have shown that two of these solutions with $(l_1, l_2) =  (3, 2), (4, 3)$ may describe a small enough (e.g. zero) variation of the effective gravitational constant $G$ (in Jordan frame)  for certain chosen values of  $\Lambda$.

 {\bf Acknowledgments}

The publication was prepared with the support of the ``RUDN University Program 5-100''.
It was also partially supported by the  Russian Foundation for Basic Research,  grant  Nr. 16-02-00602.


\end{document}